\documentstyle[preprint,aps]{revtex}
\title{Microscopic calculation of in-medium proton-proton
cross sections}
\author{G. Q. Li and R. Machleidt}
\address{Department of Physics, University of Idaho,
Moscow, ID 83843, U.S.A.}
\date{\today}
\begin{document}
\maketitle

\begin{abstract}
We derive  in-medium {\it proton-proton} cross sections
in a microscopic model based upon the Bonn nucleon-nucleon
potential and the Dirac-Brueckner approach for nuclear matter.
We demonstrate the difference between proton-proton
and neutron-proton
cross sections
and point out the need to distinguish carefully between the
two cases. We also find substantial differences between our
in-medium cross sections and phenomenological parametrizations
that are commonly used in heavy-ion reactions.
\end{abstract}

The density and/or temperature dependence of hadronic systems
is an interesting topic in nuclear physics.
Experimentally, nucleus-nucleus collisions at intermediate energies
provide a unique opportunity to form  a piece of hot nuclear
matter in the laboratory
with a density up to 2-3$\rho _0$ (with $\rho _0$, in the range of 0.15 to
0.19 fm$^{-3}$,
the saturation density of normal nuclear matter; in this paper we
use $\rho _0$=0.18 fm$^{-3}$) \cite{stock,grei}.
It is  thus  possible to study the properties of hadrons in hot
and dense media. Since this piece of dense nuclear matter exists only
for  a very short  time (typically $10^{-23}$
-$10^{-22}$ s), it is necessary
to use transport models to simulate the entire collision  process and
to deduce the properties of the intermediate stage from the known initial
conditions and  the final-state observables. At intermediate energies,
both
the mean field and the two-body collisions play an equally important role
in the dynamical evolution of the colliding system; they have to be
taken into account in the transport models on an equal footing,
together with a proper treatment of the Pauli blocking for the
in-medium two-body collisions. The Boltzmann-Uehling-Uhlenbeck (BUU)
equation \cite{bert1,mosel}
and quantum molecular dynamics (QMD)
\cite{aich3,aich4}, as well as their relativistic extensions
(RBUU and RQMD),
are promising transport models for
the description of intermediate-energy heavy-ion reactions.

Starting from the bare nucleon-nucleon (NN) interaction,
in-medium NN cross
sections have been calculated using relativistic \cite{mal1,li1} as
well as nonrelativistic \cite{fae1,fae2} Brueckner theory.
In Ref. \cite{li1}, we derived microscopically the in-medium
{\it neutron-proton} ($np$) cross sections. Our derivation was based on
the Bonn meson-exchange model for the NN
interaction \cite{mach1,mach2} and the Dirac-Brueckner approach
\cite{mach2,brock,li2} for nuclear matter.
We found that our
microscopic in-medium $np$ cross sections deviate substantially
from the phenomenological parametrization by Cugnon {\it et al}.
\cite{bert1,cugn1} which is often used in transport model calculations.

In this Brief Report, we present now our microscopic results for in-medium
{\it proton-proton} ($pp$) cross sections.
We note that the Cugnon parametrization of NN cross sections
is, in fact, a fit of the free-space $pp$ data; i.~e., no difference
is made between $np$ and $pp$ scattering.
However, since there are well-known differences between $pp$ and $np$
scattering,
one should carefully distinguish between $pp$ and $np$ cross sections.

Proton-proton scattering occurs only in states of total isospin
T=1, while $np$ exists for T=0 and 1.
This fact is responsible for the characteristic differences in the shapes
of $pp$ and $np$ differential cross sections.
This is the most crucial difference between $pp$ and $np$
and should by no means be ignored.
Furthermore, there is the Coulomb force which is involved
in $pp$ but not in $np$. Finally, in the $^1S_0$ state,
the strength of the strong interaction
shows a small difference between $pp$ and $np$
which is known as charge-independence
breaking (CIB).
However, this is a very small effect and totally negligible in our
present considerations:
the in-medium effects are by an order of magnitude larger
than CIB.

In general, in transport models such as BUU and QMD,
the electromagnetic
effects between nucleons,
mainly the Coulomb interaction, are treated separately.
So, what is needed are the in-medium $pp$
cross sections due to the strong force only.
Therefore,
we calculate in this paper the $pp$ cross sections without
the Coulomb force.
Then, the main difference between
$pp$ and $np$ cross sections is due to the fact that in the former case
only the T=1 NN channels are included while in the latter case
all T=0 and T=1 states are taken into account.
We note that our $pp$ cross sections can also be used as neutron-neutron
($nn$) cross sections, since we neglect electromagnetic effects anyhow
and the small charge-symmentry breaking, i.~e., the small difference
between the $pp$ and $nn$ strong force is totally negligible here
(cf.\ our remark, above, concerning CIB).

In this paper, we apply exactly the the same methods as in our
earlier (and more detailed) paper~\cite{li1} about $np$ cross sections to which
we refer the interested reader for details.
It is therefore sufficient to just sketch our method briefly here.
We start from the
relativistic one-boson-exchange (OBE) Bonn potential \cite{mach2} which
describes the two-nucleon system below 300 MeV
accurately. This potential is used in  (relativistic)
Dirac-Brueckner calculation for nuclear matter, in which also
the effective nucleon scalar and vector fields (the mean field)
are determined. With this
nucleon mean field and the Lorentz-boosted Pauli projector, we
solve the in-medium Thompson equation (relativistic
Bethe-Goldstone equation) to determine the $\tilde G$-matrix,
from which the in-medium NN cross sections are calculted
by identifying the $\tilde{G}$-matrix with the in-medium $K$-matrix.
As in Ref. \cite{li1}, we present our results in terms
the kinetic energy
of the incident nucleon in the  ``laboratory system" ($E_{lab}$) in which the
second nucleon is at rest.  All results shown in this paper
are obtained by using the Bonn A potential~\cite{mach2} for the bare
nuclear force;
in Ref.~\cite{li1} we have shown that the dependence of our
results on the particular model for the nuclear force
is very small (as long
as the model is quantitative and relativistic).

In Fig.~1, we show the differential cross section at
$E_{lab}$=50 (a) and 200 MeV (b) for three different
densities [$\rho=0$ (solid curves),
$\rho=\rho _0$ (dashed curves) and $\rho=2\rho _0$ (dotted curves)].
At 50 MeV, the in-medium differential cross section decreases with
increasing density. At 250 MeV, it decreases when going from $\rho=0$
to $\rho=\rho _0$ and then increases. We observed a similar behavior
in $np$~\cite{li1}.
The reason for this is that with increasing energy, the higher
partial waves become more important which are less influenced
by medium effects.
As in the case of the
$np$ differential cross sections~\cite{li1}, we have prepared
a data file, containing the $pp$ differential cross sections
as a function of angle, for a number of energies and densities. From this
data file, the $pp$ differential cross sections
for any density between 0 and 3$\rho _0$ and
any energy between 0 and 300 MeV can be obtained
with sufficient accuracy by interpolation. This data file is
available from the authors upon request.

In Fig.~2,  we compare the $pp$ differential cross section with the
$np$ one at $E_{lab}$=100 MeV and $\rho$=$\rho _0$.
Clearly there are differences between $pp$ and $np$.
The $pp$ differential cross section is almost
isotropic and has the symmetry of $d\sigma /d\Omega (\theta )=d\sigma
/d\Omega (\pi -\theta )$, while the $np$ differential cross section
is highly anisotropic and has a profound peak at backward angles.
This difference is mainly due to the fact that the T=0  states do not
contribute to $pp$.
In summary, Fig.~2 demonstrates clearly that one should distinguish
carefully between $pp$ and $np$ cross sections.

In Fig.~3, we shown the $pp$ {\it total} cross sections as a function of
the incident
energy, at $\rho$=0 (solid curves), (1/2)$\rho _0$ (dashed curves) and
(3/2)$\rho _0$ (dotted curves). The symbols represent the
exact results of our microscopic
calculation, while the curves are fits in terms of a simple and
practical
parametrization of our results:
\begin{equation}
\sigma _{pp}(E_{lab},\rho )=(23.5+0.0256(18.2-E_{lab}^{0.5})^{4.0})
{1.0+0.1667E_{lab}^{1.05}\rho ^3\over 1.0+9.704\rho ^{1.2}}
\end{equation}
where $E_{lab}$ and $\rho$ are in the units of MeV and fm$^{-3}$,
respectively.
Generaly speaking, the in-medium $pp$ total cross sections decrease
with increasing density and energy.
For completeness, we list in Table 1 the in-medium $pp$ total cross
sections as function of energy and density for some selected values.

Finally in Fig.~4, we compare the $pp$ total cross section with the $np$
one at $\rho =\rho _0$ (a) and (3/2)$\rho _0$ (b).
Also shown is the description by the Cugnon
parametrization~\cite{cugn1}.
Notice that at $\rho$=0 (Fig. 4a), our results for the $pp$ total
cross section is very close to the one by Cugnon {\it et al.}
This makes sense since the Cugnon parametrization
is a fit of the Coulomb subtracted free-space $pp$ scattering data.
At this point, we note that,
since the in-medium $pp$ cross section is always smaller
than the free one (see Fig.~3), the Cugnon parametrization
overestimates the in-medium NN cross sections.
Fig.~4a clearly demonstrates the difference between $pp$ and $np$
total cross sections. The $np$ cross sections are much larger than the
$pp$ ones, especially at low energies and densities.
At finite densities, this difference
is reduced, since the $^3S_1$ amplitude, which contributes only in $np$,
is considerable quenched in the medium.
 From Fig.~4 we learn again the it is important to distinguish between
$np$ and $pp$ cross sections.

In summary, we have presented in this Brief Report predictions for
in-medium $pp$ cross sections derived in a microcopic way.
The important conclusions are:
\begin{enumerate}
\item
There is strong density dependence in the in-medium cross
sections.
Cross sections  decrease in the medium. This indicates that
a proper treatment of the density-dependence of the in-medium NN cross sections
is important.
\item
Our microscopic predictions  for free-space $pp$ cross sections
are close to the
parametrization
developed by Cugnon {\it et al}
\cite{bert1,cugn1}.
However, at finite densities which are important in transport models,
the Cugnon parametrization, which is density
independent, overestimates the cross sections.
\item
There are substantial differences between $pp$ and $np$ cross sections
(total as well as differential). This implies that one
should carefully distinguish between $pp$ and $np$ scattering
when applying NN cross sections in transport model
calculations.
\end{enumerate}

\vspace*{1cm}

Acknowledgement: This work was supported  in part by the
U.S. National Science Foundation under Grant
No.   PHY-9211607,
and by the Idaho State Board of
Education. One of the authors (GQL) gratefully acknowledges  enlightening
discussions with Prof. C. M. Ko.
\pagebreak

\pagebreak
\centerline{Table 1}
\vskip 0.3cm
Microscopic in-medium $pp$ total cross sections in units of mb
as derived in the present work
($\rho _0\equiv 0.18$ fm$^{-3}$).
\vskip 0.3cm
\begin{tabular}{lllllll}
\hline\hline
 & \multicolumn{6}{c}{$E_{lab}$ (MeV)} \\
$\rho $&~50~&~100~&~150~&~200~&~250~&~300~\\
\hline
{}~0~&~63.38~&~35.36~&~27.18~&~23.62~&~21.76~&~21.72~\\
{}~$(1/2)\rho _0$~&~40.03~&~22.84~&~16.75~&~15.59~&~16.29~&~17.28~\\
{}~$\rho _0$~&~26.37~&~17.36~&~14.73~&~14.80~&~15.33~&~16.00~\\
{}~$(3/2)\rho _0$~&~22.24~&~17.06~&~16.19~&~16.52~&~17.37~&~17.61~\\
{}~$2\rho _0$~&~18.50~&~18.60~&~21.06~&~20.61~&~20.35~&~20.10~\\
\hline\hline
\end{tabular}
\pagebreak

\begin{figure}
\caption{In-medium {\it pp} differential cross sections
at (a) 50 MeV  and (b) 200 MeV laboratory
energy. Three densities are considered: $\rho=0$ (solid line),
$\rho=\rho_0$ (dashed line), and $\rho=2\rho_0$ (dotted line).
($\rho_0\equiv 0.18$ fm$^{-3}$)}
\end{figure}

\begin{figure}
\caption{In-medium {\it pp} and $np$ differential cross sections
at 100 MeV laboratory energy for the  density
$\rho =\rho _0=0.18$ fm$^{-3}$.}
\end{figure}

\begin{figure}
\caption{In-medium {\it pp} total cross sections  as function of
incident energy for three densities. The symbols represent the results
of our exact calculations while the curves are fits of our results
in terms of the {\it ansatz} Eq.~(1).}
\end{figure}

\begin{figure}
\caption{The in-medium $pp$ and $np$ total cross sections at
(a) $\rho$=0 and (b) $\rho=(3/2)\rho _0$ as obtained in our
microscopic derivation (solid and dashed line, resp.) are compared
to the description of NN cross sections by the Cugnon parametrization.}
\end{figure}

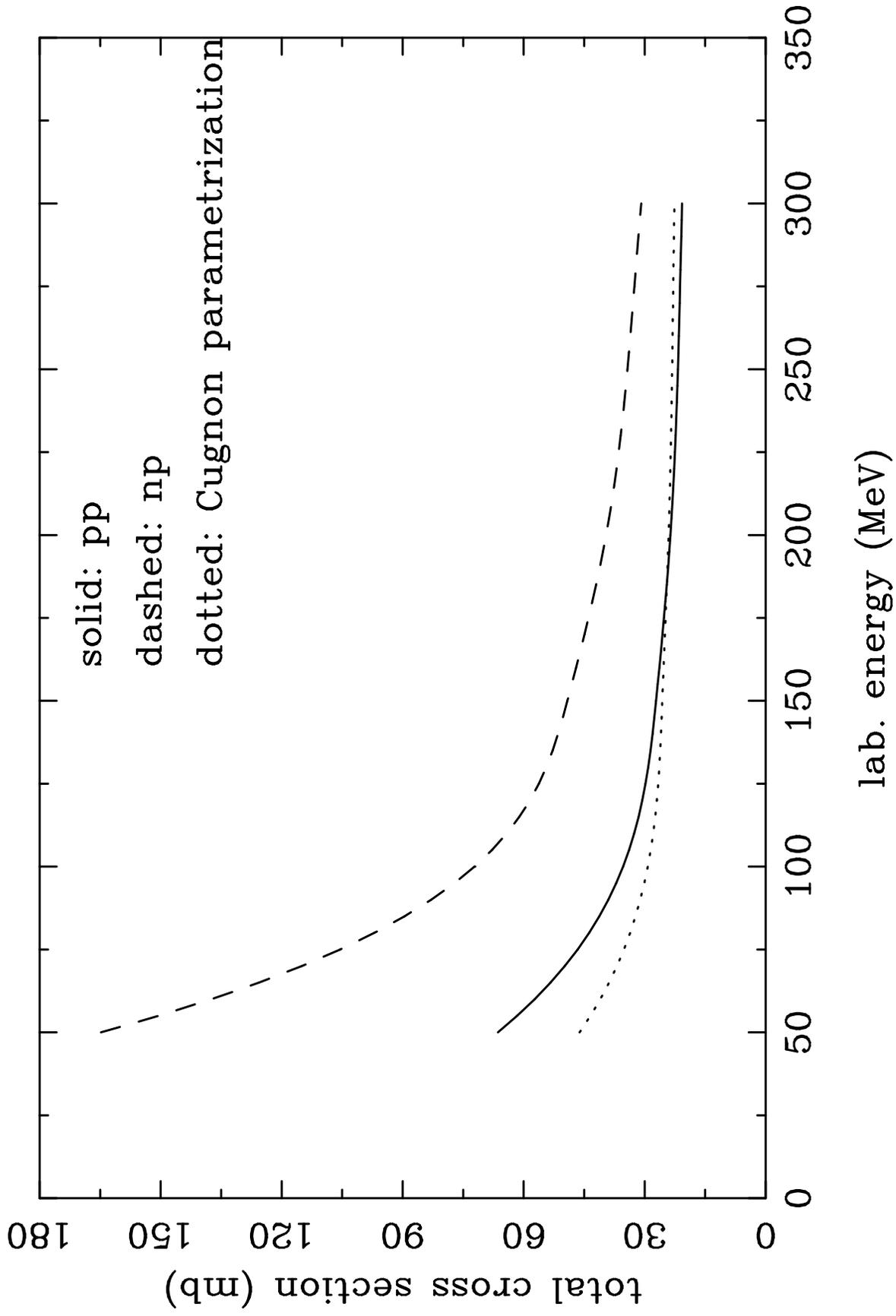
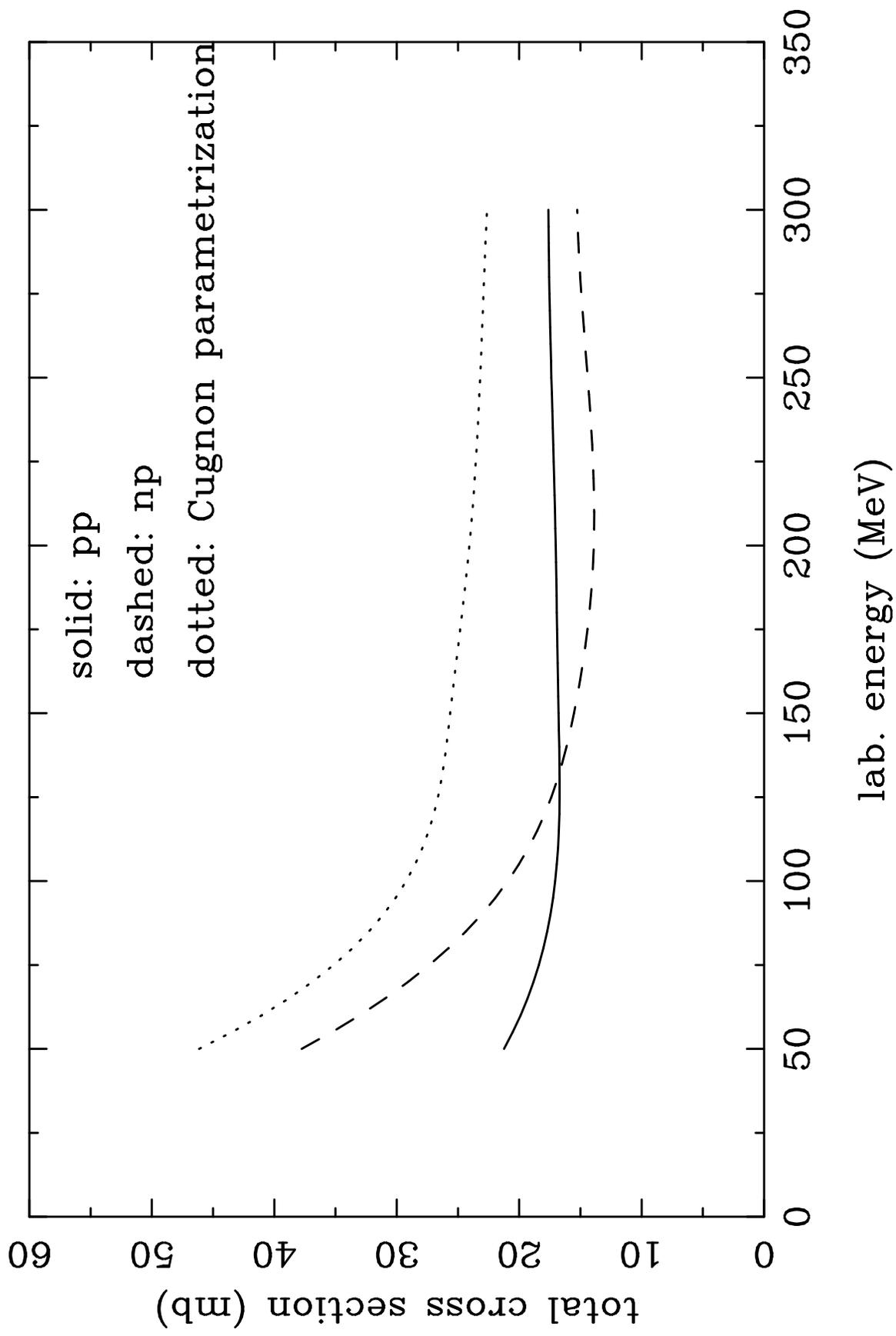
\end{document}